\documentclass[10pt,a4paper]{article}

\usepackage[cp1250]{inputenc}
\usepackage[czech]{babel}

\begin{document}

\renewcommand{\thesubsection}{\arabic{subsection}}
\pagestyle{plain}
\setcounter{page}{1}

\begin{center}
\section*{The Kerr--Schild ansatz for the Nariai spacetime and the generating conjecture}

\vspace*{6mm}

M{\small{ILAN}} \v S{\small{TEFAN\' IK}}, J{\small AN} H{\small ORSK\' Y}

{\it{Department of Theoretical Physics and Astrophysics, Faculty of Science,\\
Masaryk University, Kotl\' a\v rska 2, 611 37 Brno, Czech Republic. \\
stefanik@physics.muni.cz, horsky@physics.muni.cz}}
\end{center}

\vspace*{6ex}

\noindent PACS: 04.20.Jb, 04.40.Nr

\vspace*{6ex}

{\bf Abstract}

\vspace*{5mm}

\noindent {\small We will show that the Nariai metric, i.e. the static spherically symmetric vacuum spacetime with a cosmological constant, admits a conformally Kerr--Schild spacetime representation.  We find the vacuum solutions of the Einstein--Maxwell equations for the Nariai metric using the Horsk\' y--Mitskievich generating conjecture.}

\vspace*{3ex}

\subsection{Introduction}

Some static spherically symmetric spacetimes can be written in the form
\begin{displaymath}
ds^{2} = e^{2\alpha}dt^{2} - e^{2\beta}d{\rho}^{2} - {\rho}^{2}\left(d{\vartheta}^{2} + 
{{\sin}^{2} \vartheta}d{\varphi}^{2} \right) .
\end{displaymath}
\noindent Here $\left(t, \rho ,\vartheta , \varphi \right) $ is a coordinate system and $ \alpha , \beta $ are the functions of $\rho$ only.
There exists the Nariai metric \cite{Kra-Her80} which also describes the static and spherically symmetric solution of the Einstein equations with non zero cosmological constant $ \Lambda$. This spacetime has the form
\begin{equation}\label{1}
ds^{2} = \left(1 - \Lambda {\rho}^{2} \right)dt^{2} - \frac{1}{\left(1 - \Lambda {\rho}^{2} \right)}d{\rho}^{2} -
\frac{1}{\Lambda }\left(d{\vartheta}^{2} + {{\sin}^{2} \vartheta}d{\varphi}^{2} \right) .
\end{equation}
\noindent In the second paragraph we will show that this static and spherically symmetric spacetime with a cosmological constant can also be transformed to the conformally Kerr--Schild form \cite{Mi-Ho96}. 

There are many techniques how to obtain the solutions of the Einstein equations and the Einstein--Maxwell equations of the general relativity, i.e. \cite{Kra-Her80}, \cite{Lu83}, \cite{Hoe-Die84}. For the Nariai metric (\ref{1}) we use the Horsk\' y--Mitskievich generating conjecture originally formulated in \cite{Ho-Mi89}. It is possible to find the exact solutions of 
the Einstein--Maxwell equations if a seed metric i.e. a spacetime without an electromagnetic field, has one Killing vector at least.
This conjecture is a generalization of the fact that the timelike Killing covectors of the 
Schwarzschild and the Kerr solutions coincide (up to a constant factor) with the electromagnetic four-potentials of the 
Reissner--Nordstr\" om and the Kerr--Newman fields respectively \cite{Ca-Ku-Mi94}. In the case of the Nariai metric (\ref{1}) we will consider the Killing vectors $ {\partial}_{t} $ and $ {\partial}_{\varphi} $.

\vspace*{3ex}

\subsection{The Kerr--Schild form of the Nariai metric}

The Nariai metric (\ref{1}) can be rewritten in the form
\begin{equation}\label{2a}
ds^{2} = \frac{1}{\Lambda {\rho}^{2}}\left[e^{2\alpha}dt^{2} - e^{2\beta}d{\rho}^{2} - {\rho}^{2}
\left(d{\vartheta}^{2} + {\sin}^{2}\vartheta d{\varphi}^{2} \right) \right] ,
\end{equation}
\noindent where
\begin{equation}\label{2b}
e^{2\alpha} = \Lambda {\rho}^{2}\left(1 - \Lambda {\rho}^{2} \right), \qquad \qquad
e^{2\beta } = \frac{\Lambda {\rho}^{2}}{\left(1 - \Lambda {\rho}^{2} \right)} .
\end{equation}
\noindent Let us prove that (\ref{1}) or (\ref{2a}) can be written in the conformally Kerr--Schild metric representation
\begin{equation}\label{3}
ds^{2} = e^{2U}\left[ds^{2}_{M} - 2H\left(d\tau \pm dr \right)^{2} \right] ,
\end{equation}
\noindent where $ U = U\left(\tau, r \right)$, $ H = H\left(\tau, r \right)$ and $ds^{2}_{M}$ is the Minkowski metric (not necessarily in the Cartesian coordinates). 

\noindent The calculations follow directly from the pattern presented in \cite{Mi-Ho96}. Let us consider the transformation
\begin{equation}\label{3a}
dt = d\tau + f(\rho)d\rho, \qquad \qquad \rho = {\rho}(r) ,
\end{equation}
which transforms the spacetime (\ref{1}) or (\ref{2a}) to the form (\ref{3}). In the case of the Nariai metric the function $ f(\rho)$ has the form
\begin{displaymath}
f(\rho) = - \frac{1}{\Lambda {\rho}^{2}}\cdot\frac{\left(1 - 2\Lambda {\rho}^{2}\right)}{\left(1 - \Lambda {\rho}^{2}\right)} ,
\end{displaymath}
\noindent and the term from the paper \cite{Mi-Ho96}
\begin{displaymath}
ds^{2} = \frac{{\rho}^{2}}{r^{2}}\left[ds^{2}_{M} - 2H\left(d\tau \pm dr \right)^{2}\right] ,
\end{displaymath}
\noindent takes the following form
\begin{equation}\label{4}
ds^{2} = \frac{1}{\Lambda r^{2}}\left[ds^{2}_{M} - 2H\left(d\tau \pm dr \right)^{2} \right] ,
\end{equation}
\noindent for the case of the Nariai metric. Here
\begin{equation}\label{5}
2H = 1 - \frac{r^{2}}{{\rho}^{2}}e^{2\alpha } ,
\end{equation}
\noindent and $\alpha = \alpha({\rho}(r))$. The function ${\rho}(r)$ is found by integration of the relation
\begin{equation}\label{6}
\frac{d\rho}{dr} = \frac{{\rho}^{2}}{r^{2}}e^{-\left(\alpha + \beta  \right)} ,
\end{equation}
\noindent where the functions $\alpha$ and $\beta$ are given by (\ref{2b}). Substituting for $\alpha$ and $\beta$ to (\ref{6}) and after the integration we get
\begin{equation}\label{7}
\rho = - \frac{1}{\Lambda r} .
\end{equation}
\noindent Therefore the relationship (\ref{5}) is equal to
\begin{equation}\label{8}
2H = 2\left(1 - \frac{\Lambda r^{2}}{2} \right) .
\end{equation}
\noindent In such a way we have obtained the final form of (\ref{3})
\begin{equation}\label{9}
ds^{2} = \frac{1}{\Lambda r^{2}}\left[ds^{2}_{M} - 2\left(1 - \frac{\Lambda r^{2}}{2} \right)\left(d\tau - dr \right)^{2} \right] .
\end{equation}
We have shown that the static and spherically symmetric spacetime with a cosmological constant (\ref{1}) can also be transformed to the conformally Kerr--Schild form \cite{Mi-Ho96}. 
We have completed, therefore, the assertion that it is possible to write every spherically symmetric solution of the Einstein equations in the conformally Kerr--Schild form (\ref{3}).

\vspace*{3ex}

\subsection{The application of the Horsk\' y--Mitskievich conjecture to the Nariai metric}

In 1989 the Horsk\' y--Mitskievich conjecture was formulated which states that \cite{Ho-Mi89} {\it a test electromagnetic field having a potential proportional to the Killing vector of a seed vacuum gravitational field (up to a constant factor), is simultaneously an exact electromagnetic potential of the self consistent Einstein--Maxwell problem where the metric is 
stationary or static and it reduces to the mentioned seed metric when the parameter characterizing the charge of the
electromagnetic field source vanishes}.

This generating conjecture we will apply to the Nariai spacetime and to obtain a solution of the Einstein--Maxwell field equations. 

\noindent The Nariai metric $g^{\rm{(seed)}}_{mn}$ (\ref{1}) has two Killing vectors
\begin{equation}\label{12}
T^{i} = \partial _{t} , \qquad \Phi^{i} = \partial _{\varphi } . 
\end{equation}
The corresponding covector forms are
\begin{eqnarray*}
T_{m} & = & g^{\rm{(seed)}}_{mn} T^{n}  =  \left ( 1 - \Lambda {\rho}^{2} \right ),      \\
\Phi_{p} & = & g^{\rm{(seed)}}_{pq} \Phi^{q}  =  -\frac{\sin^{2} \theta}{\Lambda},
\end{eqnarray*}
and, consequently, the 4--potentials
\begin{eqnarray}
\label{13}
A_{m} &=& \beta T_{m} = \beta \left ( 1 - \Lambda {\rho}^{2} \right ), \quad \mbox{for Killing vector} \quad \partial _{t} ,     \\
\label{14}
A_{p} &=& \beta \Phi _{p} = -\beta \frac{\sin^{2} \theta}{\Lambda}, \qquad \mbox{for Killing vector} \quad \partial _{\varphi } .
\end{eqnarray}
Here $ \beta $ is a parameter characterizing the charge of the electromagnetic field source.

\noindent The non-zero components of the electromagnetic field tensor are
\begin{eqnarray*}
F_{t \rho} &=& 2\beta \Lambda \rho , \qquad \qquad \qquad \mbox{for} \quad \partial _{t} ,  \\
F_{\theta \varphi} &=& -\frac{\beta }{\Lambda }\sin 2\theta , \qquad \qquad \mbox{for} \quad \partial _{\varphi} .
\end{eqnarray*}
Because the 4-potential (\ref{13}) is a function of $\rho$-coordinate and the 4-potential (\ref{14}) is a function of $\theta$-coordinate  
the charged metrics of the self consistent Einstein--Maxwell problem are supposed to have the~form
\begin{equation}\label{15}
ds^{2}_{new} = e^{2A(\rho)}{dt}^{2} - e^{2B(\rho)}{d\rho}^{2} - e^{2C(\rho)}{d\theta}^{2}  - e^{2D(\rho)}{d\varphi}^{2} ,
\end{equation}
for the 4-potential (\ref{13}) and 
\begin{equation}\label{16}
ds^{2}_{new} = e^{2A(\theta)}{dt}^{2} - e^{2B(\theta)}{d\rho}^{2} - e^{2C(\theta)}{d\theta}^{2}  - e^{2D(\theta)}{d\varphi}^{2} , 
\end{equation}
for the 4-potential (\ref{14}).
From the Maxwell vacuum equations we obtain the following condition for the spacetime  (\ref{15}) 
\begin{equation}\label{17}
C' + D' - A' -B' +\frac{1}{\rho} = 0,
\end{equation}
and for the metric  (\ref{16}) 
\begin{equation}\label{18}
\dot A + \dot B - \dot C -\dot D + \frac{2\cos 2\theta }{\sin 2\theta} = 0,
\end{equation}

\noindent where a prime means the derivative with respect to $\rho$ and a dot means the derivative with respect to $\theta$.

\noindent Let us write the Einstein equations in the form 
\begin{displaymath}
R_{ik}  - \frac{1}{2}Rg_{ik} - \Lambda g_{ik}= 8\pi T_{ik} .
\end{displaymath}
\noindent The Ricci tensor $R_{ik}$, the Ricci scalar $R$ and the energy--momentum tensor $T_{ik}$ are defined by \cite{MTW}. 
For the metric (\ref{15}) these equations are
\begin{eqnarray}
\label{19}
C'' + C'^{2} + D'' + D'^{2} - B'\left(C' + D' \right) + D'C' + \Lambda e^{2B} &=& -4\left(\beta \Lambda \rho  \right)^{2}e^{-2A} ,\\
\label{20}
C'A' + D'A' +D'C' + \Lambda e^{2B} &=& -4\left(\beta \Lambda \rho  \right)^{2}e^{-2A} ,\\
\label{21}
A'' + A'^{2} + D'' + D'^{2} - B'\left(A' + D' \right) +A'D' + \Lambda e^{2B} &=& 4\left(\beta \Lambda \rho \right)^{2}e^{-2A} ,\\
\label{22}
A'' + A'^{2} + C'' + C'^{2} - B'\left(A' + C' \right) +A'C' + \Lambda e^{2B} &=& 4\left(\beta \Lambda \rho \right)^{2}e^{-2A} .
\end{eqnarray}
For the Einstein equations with respect to the spacetime (\ref{16}) we obtain 
\begin{eqnarray}
\label{23}
\ddot B + {\dot B}^{2} + \ddot D + {\dot D}^{2} - \dot C\left(\dot B + \dot D \right) + \dot D\dot B + \Lambda e^{2C}     &=&
 -\frac{1}{2}\left(\frac{\beta }{\Lambda } \right)^{2}\left(1 - \cos 4\theta  \right)e^{-2D} ,        \\
\label{24}
\ddot A + {\dot A}^{2} + \ddot D + {\dot D}^{2} - \dot C\left(\dot A + \dot D \right) + \dot D\dot A + \Lambda e^{2C}     &=&
 -\frac{1}{2}\left(\frac{\beta }{\Lambda } \right)^{2}\left(1 - \cos 4\theta  \right)e^{-2D} , \\
\label{25}
\dot B\dot A + \dot D\dot A + \dot D\dot B + \Lambda e^{2C}    &=&  \frac{1}{2}\left(\frac{\beta }{\Lambda } \right)^{2}\left(1 - \cos 4\theta  \right)e^{-2D} ,       \\
\label{26}
 \ddot B + {\dot B}^{2} + \ddot A + {\dot A}^{2} - \dot C\left(\dot B + \dot A \right) + \dot A\dot B + \Lambda e^{2C}     &=&
\frac{1}{2}\left(\frac{\beta }{\Lambda } \right)^{2}\left(1 - \cos 4\theta  \right)e^{-2D} .
\end{eqnarray}

\noindent The system of the equations (\ref{17}) and (\ref{19} - \ref{22}) should determine $A\left(\rho \right), B\left(\rho \right), C\left(\rho \right)$ and  $D\left(\rho \right)$. 
We have thus to find the four functions from a system of the five equations. Fortunately, this system is not independent therefore we can easily obtain the solutions. The four unknown functions are
\begin{eqnarray}
\label{27}
e^{2C\left(\rho \right)}        &=& \left[c_{2}\left(\rho ^{2} -2c_{1}\right) \right]^{-2} ,   \\
\label{28}
e^{2D\left(\rho \right)}        &=&  \left[c_{2}\left(\rho ^{2} -2c_{1}\right) \right]^{-2} ,    \\
\label{29}
e^{2A\left(\rho \right)}        &=& \frac{\left(\beta \Lambda \rho \right)^{2}\left(\rho ^{6} - 24c_{1}^{2}\rho ^{2} +64c_{1}^{3}\right)}{\left(\rho ^{2} - 2c_{1} \right)^{2}}
- \frac{c_{3}}{\left(\rho ^{2} - 2c_{1} \right)^{2}} - \frac{c_{4}\rho ^{2}\left(\rho ^{4} - 6c_{1}\rho ^{2} + 12c_{1}^{2}\right)}{\left(\rho ^{2} - 2c_{1} \right)^{2}} ,         \\
\label{30}
\Lambda e^{2B\left(\rho \right)}        &=& -\left[4\beta ^{2}\Lambda ^{2}\rho ^{2}e^{-2A\left(\rho  \right)} + 2C'\left(\rho  \right)A'\left(\rho  \right) + C'\left(\rho  \right)^{2}\right] ,
\end{eqnarray}
where $c_{1}, c_{2}, c_{3}$ and $c_{4}$ are integration constants.

\noindent For the system of the equations (\ref{18}) and (\ref{23} - \ref{26}) we get using the same way the solutions
\begin{eqnarray}
\label{31}
e^{2A\left(\theta \right)}        &=& \left[ k_{2} \left( \cos 2 \theta + 2 k_{1} \right) \right]^{-2} ,        \\
\label{32}
e^{2B\left(\theta \right)}        &=&  \left[ k_{2} \left( \cos 2 \theta + 2 k_{1} \right) \right]^{-2} ,        \\
\label{33}
e^{2D\left(\theta \right)}        &=&  k_{4}\cos ^{2}\theta \left[1 - \frac{1}{\left( \cos 2 \theta + 2 k_{1} \right)}
+\frac{1-6k_{1}+8k_{1}^{2}+4k_{1}\cos ^{2}\theta}{\left( \cos 2 \theta + 2 k_{1} \right)^{2}} \right]        \nonumber \\
   & & + k_{3}\sin ^{2}\theta \left[1+\frac{2\left(2k_{1}+1 \right)}{\left( \cos 2 \theta + 2 k_{1} \right)} 
+\frac{2\sin ^{2}\theta \left(1+2k_{1} \right)}{\left( \cos 2 \theta + 2 k_{1} \right)^{2}}  \right]                   \nonumber \\
   & & - \frac{2\beta ^{2}\cos ^{4}\theta }{3\Lambda ^{2}}\left[1+\frac{2\left(2k_{1}-1 \right)}{\left( \cos 2 \theta + 2 k_{1} \right)} 
+\frac{2\cos ^{4}\theta }{\left( \cos 2 \theta + 2 k_{1} \right)^{2}}\right] ,        \\
\label{34}
\Lambda e^{2C\left(\theta \right)}        &=&  \frac{\beta ^{2}}{2\Lambda ^{2}}
\left(1-\cos 4\theta  \right) e^{-2D\left(\theta  \right)} -2 \dot A\left(\theta  \right)\dot D\left(\theta  \right)-\dot A \left(\theta \right)^{2} ,
\end{eqnarray}
where $k_{1}, k_{2}, k_{3}$ and $k_{4}$ are integration constants.
\noindent The invariants of the electromagnetic field of our solutions are
\begin{eqnarray}
\label{35}
F_{ik}F^{ik} &=& -8\frac{\beta ^{2}\Lambda ^{2}\rho ^{2}}{e^{2A\left(\rho  \right)+2B\left(\rho  \right)}} < 0 \;\qquad \mbox{for} \quad \partial _{t} ,          \\
        &        &    \nonumber      \\
\label{36}
F_{ik}F^{ik} &=& 8\frac{\beta ^{2}\sin ^{2}\theta \cos ^{2}\theta }{\Lambda ^{2}e^{2C\left(\theta  \right)+2D\left(\theta  \right)}} > 0 \qquad \mbox{for} \quad \partial _{\varphi } .
\end{eqnarray}
This means that the field is an electric type for the Killing vector $\partial _{t}$ and a magnetic type for the Killing vector $\partial _{\varphi }$. Both solutions (\ref{27}--\ref{30}) and (\ref{31}--\ref{34}) admit three Killing vectors. In the first case they are $\partial _{t }, \partial _{\theta }, \partial _{\varphi }$ and $\partial _{t }, \partial _{\rho }, \partial _{\varphi }$ for the second one. 
This is differed from the seed spacetime (\ref{1}) because we have assumed the higher symmetry of the charged solutions (\ref{15}) and (\ref{16}). 

\vspace*{3ex}

\subsection{Conclusion}

We have considered the static and spherically symmetric vacuum spacetime with the non zero cosmological constant which is called the Nariai metric (\ref{1}). For this metric we have shown that this spacetime can be transformed to the conformally Kerr--Schild form (\ref{3}). Therefore the statement that {\it all static spherically symmetric spacetimes can be described by conformally Kerr--Schild metrics} used in \cite{Mi-Ho96} really holds for all static spherically symmetric vacuum spacetimes with or without the cosmological constant.
  
For the Nariai metric (\ref{1}) we have found the corresponding spacetime with an electromagnetic field generated by the Horsk\'y --Mitskievich conjecture. 
These solutions are really given explicitly by (\ref{27}--\ref{30}) for the Killing vector $T^{i}$ and by (\ref{31}--\ref{34}) for the 
Killing vector $\Phi ^{i}$. Each of these solutions already has three Killing vector fields, compared to two Killing vectors for the seed metric (\ref{1}).

\vspace*{3ex}

\subsection*{Acknowledgements}

We would like to thank Professor P. C. Aichelburg, Professor R. Beig and Professor H. Urbantke from the University in Vienna for their initial help. We are grateful to Professor N. V. Mitskievich for fruitful discussions.

\end{document}